\documentclass{optica-article}

\journal{opticajournal} 

\articletype{Research Article}

\usepackage{lineno}

\usepackage{glossaries-extra}
\setabbreviationstyle[acronym]{long-short}

\newacronym{fso}{FSO}{Free-Space Optical}
\newacronym{rf}{RF}{Radio Frequency}
\newacronym{qkd}{QKD}{Quantum Key Distribution}
\newacronym{smf}{SMF}{Single-Mode Fiber}
\newacronym{ce}{CE}{Coupling Efficiency}
\newacronym{ao}{AO}{Adaptive Optics}
\newacronym{fsm}{FSM}{Fast-Steering Mirror}
\newacronym{dm}{DM}{Deformable Mirror}
\newacronym{dl}{DL}{Deformable Lens}
\newacronym{fsp}{FSP}{Fast-Steering Prism}
\newacronym{mal}{MAL}{Multi-Actuator deformable Lens}
\newacronym{swir}{SWIR}{Short-Wave InfraRed}
\newacronym{adc}{ADC}{Analog-to-Digital Converter}
\newacronym{wfs}{WFS}{Wave-Front Sensor}

\begin{document}

\title{Multi-actuator lens systems for turbulence correction in free-space optical communications}

\author{Matteo Schiavon,\authormark{1,*} Antonio Vanzo,\authormark{2,*} Kevin Campaci,\authormark{2} Valentina Marulanda Acosta,\authormark{1,3} and Stefano Bonora\authormark{2}}

\address{\authormark{1}Sorbonne Universit\'e, CNRS, LIP6, F-75005 Paris, France\\
\authormark{2}CNR, Istituto di Fotonica e Nanotecnologie, 35131 Padova, Italy\\
\authormark{3}DOTA, ONERA, Universit\'e Paris Saclay, F-92322, Ch\^atillon, France}

\email{\authormark{*}matteo.schiavon@lip6.fr} 


\begin{abstract*}
The implementation of efficient free-space channels is fundamental for both classical and quantum Free-Space Optical (FSO) communication.
This can be challenging for fibre-coupled receivers, due to the time variant inhomogeneity of the refractive index that can cause strong fluctuations in the power coupled into the Single-Mode Fiber (SMF), and requires the use of Adaptive Optics (AO) systems to correct the atmospheric induced aberrations.
In this work, we present two adaptive optic systems, one using a Fast-Steering Prism (FSP) for the correction of tip-tilt and a second one based on a Multi-Actuator deformable Lens (MAL), capable of correcting up to the third order of Zernike's polynomials.
We test both systems at telecom wavelength both with artificial turbulence in the laboratory and on a free-space channel, demonstrating their effectiveness in increasing the fibre coupling efficiency.
\end{abstract*}

\section{Introduction}

\gls{fso} communication presents some undeniable advantages with respect to \gls{rf} links for the implementation of point-to-point communication, in terms of both transmission efficiency and security.
The high frequency of optical beams leads to high speed modulation, allowing higher data rates, and low beam divergence, which maximizes the power intercepted by the legitimate receiver, increasing both transmission efficiency and security.
The use of optical frequencies can also been exploited for the use of \gls{qkd}, which allows the exchange of secret keys with unconditional security~\cite{Scarani2009}.

The importance of high speed communication justifies the effort for the development of new devices and techniques for pushing the communication towards its fundamental limits~\cite{Jrgensen2022}.
This research is mainly focused on fibre systems, which means that the most performing communication devices are based on \gls{smf} technology and work at telecom wavelength.
For the cases where the installation of optical fibres is not possible, for geographic or economic reasons, \gls{fso} communication represents a good alternative~\cite{Jahid2022}.
However, the free space channel is affected by atmospheric turbulence, which severely decreases the \gls{ce} into the \gls{smf}.
This requires the use of \gls{ao} for partially compensating the effects of the turbulence and increasing the amount of light coupled into the fibre~\cite{Wu2010,Li2014,Liu2014,Chen2015,Li2016,Wang2018,Walsh2022}.

The working principle of \gls{ao} systems consists in measuring wavefront aberrations and correcting it using active optical elements~\cite{Tyson2010}.
The typical \gls{ao} system works using reflecting elements, such as a \gls{fsm} for the correction of the first order of the turbulence (tip-tilt) and a \gls{dm} for the correction of higher orders.
The use of transmitting elements, on the other hand, could open the way to more compact systems, which represents a very promising feature for \gls{fso} systems.
The use of a \gls{mal} for coupling into a \gls{smf} has already been demonstrated in visible wavelengths~\cite{Moosavi2019}.
However, the use of such devices for \gls{fso} communication requires a study of its performance at telecom wavelength, in which high-speed telecommunication devices work.

In this article, we investigate the performance of two different \gls{ao} systems, based on transmitting devices, which work at telecom wavelength.
The first system is based on a \gls{fsp} and can correct the first order of the turbulence (tip-tilt).
The second system uses a \gls{mal}, which is capable of correcting up to the fourth order of the turbulence.
Both systems are tested using induced atmospheric turbulence in the laboratory, and the performance of the second one is also assessed on a 100-m ground-to-ground free-space channel.

\section{Materials and methods}

\subsection{The fast-steering prism}
\label{sec:fsp}

\begin{figure}[!h]
    \centering
    \includegraphics[width=0.8\textwidth]{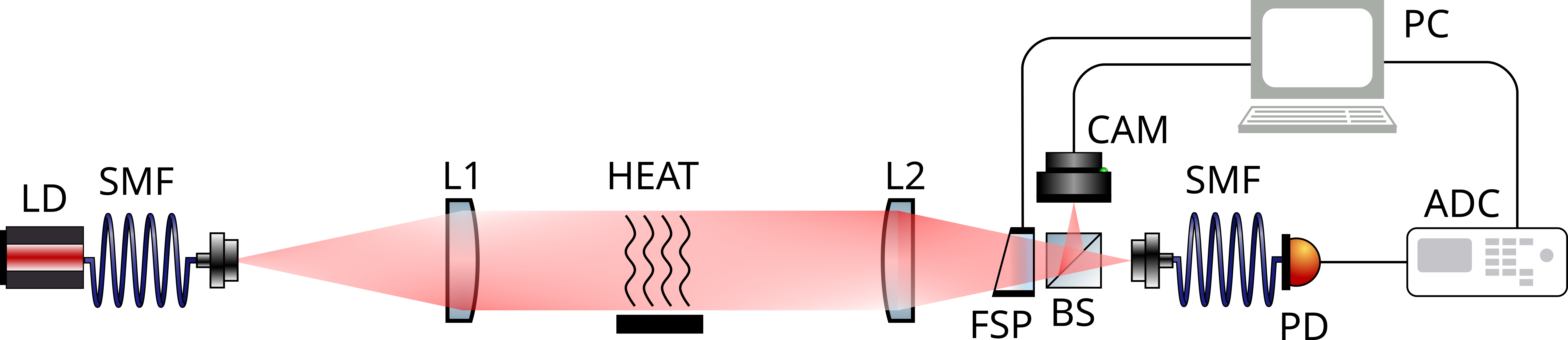}
    \caption{Optical setup used for the test of the \gls{fsp} in the laboratory.}
    \label{fig:fsp}
\end{figure}

The optical setup used to test the performance of the \gls{fsp} is shown in Figure~\ref{fig:fsp}.

A laser diode (LD) at telecom wavelength $\lambda = 1550 \, nm$ (MMC-GASPFP-V Mini Media Converter), coupled to a \gls{smf} of numerical aperture $NA = 0.14$ and mode field diameter $MFD = 10.4 \, \mu m$, is collimated into a beam of diameter $50.8 \, mm$ (Rayleigh length $z_R > 1.6 \, km$) using a lens of focal length $300 \, mm$ (L1).
The beam is propagated on the optical bench for a distance of around $2 \, m$, with an electric heater (HEAT) placed in the middle of the path to generate the turbulence.
The beam is collected by the receiving telescope, consisting of a lens of diameter $50.8 \, mm$ and focal length $300 \, mm$, which focuses it into a \gls{smf} ($NA = 0.14$, $MFD = 10.4 \, \mu m$).
The \gls{fsp} is placed between the lens L2 and the fibre, before an unbalanced beam-splitter (BS) that sends $70\%$ of the light to the fibre and $30\%$ to a \gls{swir} camera (CAM).
The camera (FirstLight C-RED 2 Lite) has a full-frame rate of $600 \, FPS$ and is linked to the control computer (PC) using a USB 3.0 connection.
The images are processed by the Dynamics Optics PhotonLoop software ~\cite{Mocci2018}, that calculates the position of the centroid and uses it as an error signal to calculate the voltage to apply to the \gls{fsp} through a Piezo Stack driver (Dynamic Optics srl).
The \gls{fsp}, developed at the CNR-IFN laboratory ~\cite{Vanzo2023}, is composed of two glass windows with a dielectric gel in-between. Glass surfaces are uncoated for a total transmittance of about $85\%$.
The first window is fixed, while the second one can be tilted by $3$ amplified piezoelectric actuators.
The prism has a clear aperture of diameter $23 \, mm$ and it can reach up to $\pm 2.2 \, mrad$ of optical tilt.
The head of the fibre is mounted on a manual x-y stage and its position along the optical axis is controlled by a motorized translation stage.
The power collected by the fibre is measured by a photodiode (Thorlabs DET01CFC, PD), whose output, on a $100 \, k\Omega$ load, is read using a National Instruments USB-6009 \gls{adc}.
The \gls{ce} is measured as the ratio between the power read by the photodiode and the power before the fibre, measured using a powermeter.

\subsection{The multi-actuator lens}

\begin{figure}[!h]
    \centering
    \includegraphics[width=1\textwidth]{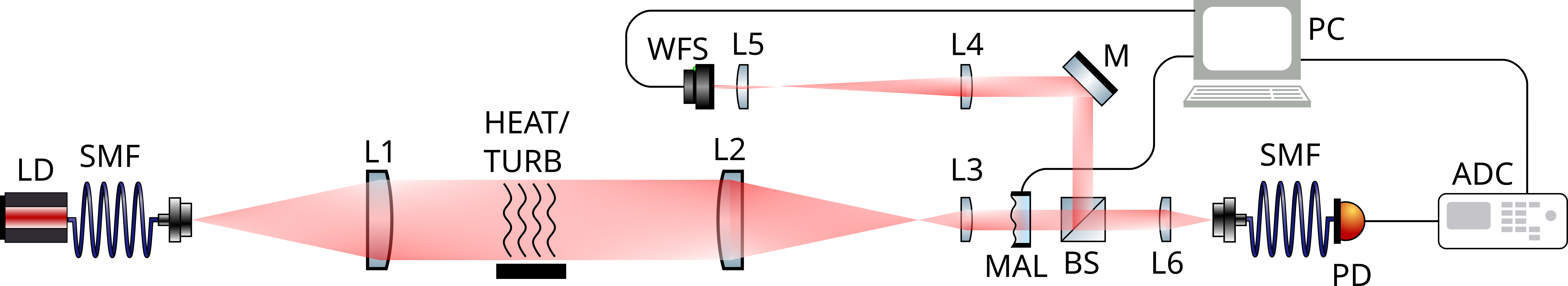}
    \caption{Optical setup used for the test of the \gls{mal} in the laboratory.}
    \label{fig:mal}
\end{figure}

The \gls{mal} is tested using the optical setup shown in Figure~\ref{fig:mal}.

The transmitting system used for testing the performance of the \gls{mal} is equal to the one described in Section~\ref{sec:fsp}, producing a collimated beam of diameter $50.8 \, mm$ that propagates on top of an electric heater.
The receiving system is a Keplerian telescope, composed of a first lens of diameter $50.8 \, mm$ and focal lengths $300 \, mm$ (L2) and a second lens of focal length $60 \, mm$ (L3), that reduces the diameter of the beam to $10.2 \, mm$.
The \gls{mal} is positioned at the point conjugate to L2, $74 \, mm$ behind L3.
After the \gls{mal}, an unbalanced beam-splitter sends $10\%$ of the light towards the \gls{wfs} and $90\%$ of the light towards the \gls{smf}.
The \gls{wfs} is a Shack-Hartmann built using a \gls{swir} camera (FirstLight C-RED 2 Lite), with full-frame rate $600 \, FPS$, and an array of 6 by 6 microlenses with a focal length of about $5.6 \, mm$ and a pitch of $250 \, \mu m$.
The \gls{wfs} is placed after a second Keplerian telescope with lenses of focal length $300 \, mm$ (L4) and $40 \, mm$ (L5), in the point conjugated with the \gls{mal}.
The beam is focused into the \gls{smf} with a lens of focal length $70 \, mm$ (L6).
Similarly to the system described in Section~\ref{sec:fsp}, the head of the fibre is mounted on a manual x-y stage and the position on the optical axis is controlled by a motorized stage.
The measurement of the coupled power and the calculation of the \gls{ce} is performed as described in Section~\ref{sec:fsp}.

The \gls{mal} ~\cite{Quintavalla2020}, produced by Dynamic Optics, is composed of two thin glass windows of thickness $150 \, \mu m$, with a piezoelectric actuator ring mounted upon them. The clear aperture of the MAL is 10mm.
The space between the windows is filled with a transparent gel.
The first window is used to generate defocus and astigmatism while the second one generates coma and secondary astigmatism.
Both rings are divided into sectors that can be actuated independently.
The piezo rings are glued to the windows and act as a bimorph actuator, bending the window at the application of a voltage.
The actuators are controlled by a high voltage driver (Dynamic Optics PZTMini), that can provide up to $125 \, V$.
The driver is controlled by the Dynamic Optics PhotonLoop software, that analyzes the wavefront images read by the \gls{wfs} at a rate of $600 \, FPS$ and calculates the output voltages that are sent to the \gls{mal}. \\

\begin{figure}[!h]
    \centering
    \includegraphics[width=1\textwidth]{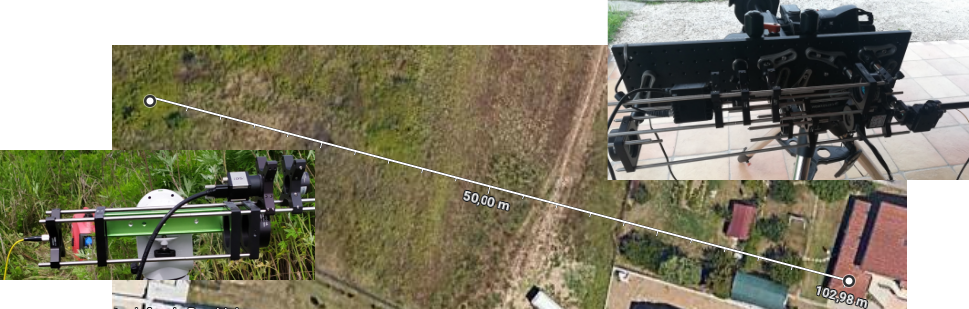}
    \caption{The ground-to-ground channel in which the \gls{mal} was tested, with a photo of the transmitting (left) and of the receiving (right) system.}
    \label{fig:g2g}
\end{figure}

The \gls{mal} was also tested on the $100 \, m$ ground-to-ground channel shown in Figure~\ref{fig:g2g}.
Both the transmitter and the receiver use the setup shown in Figure~\ref{fig:mal}.
The transmitter is mounted on a motorized tripod (Sky-Watcher SUNGT Solar Altazimuth Mount) and a camera (IDS UI-3060CP-M-GL), combined with an achromatic lens with a focal length of $100 \, mm$, is used to provide a coarse pointing system towards the receiver, which is also mounted on a motorized tripod (Celestron CGX).

\section{Results}
\subsection{Laboratory experiments}

\begin{figure}[htbp]
\centering
\includegraphics[width=4.3cm]{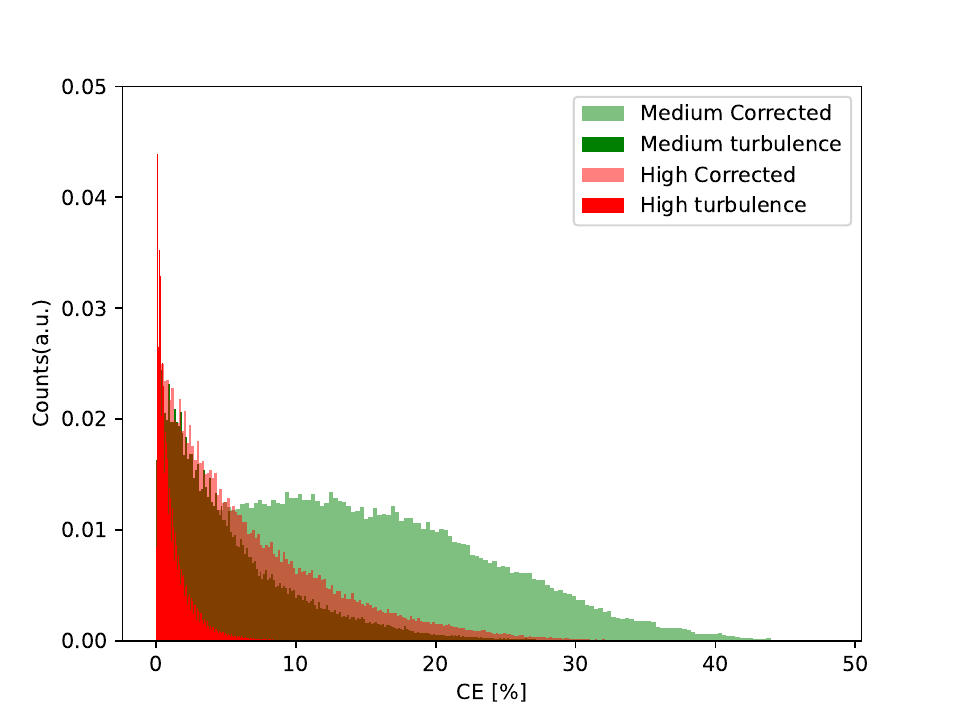}
\includegraphics[width=4.3cm]{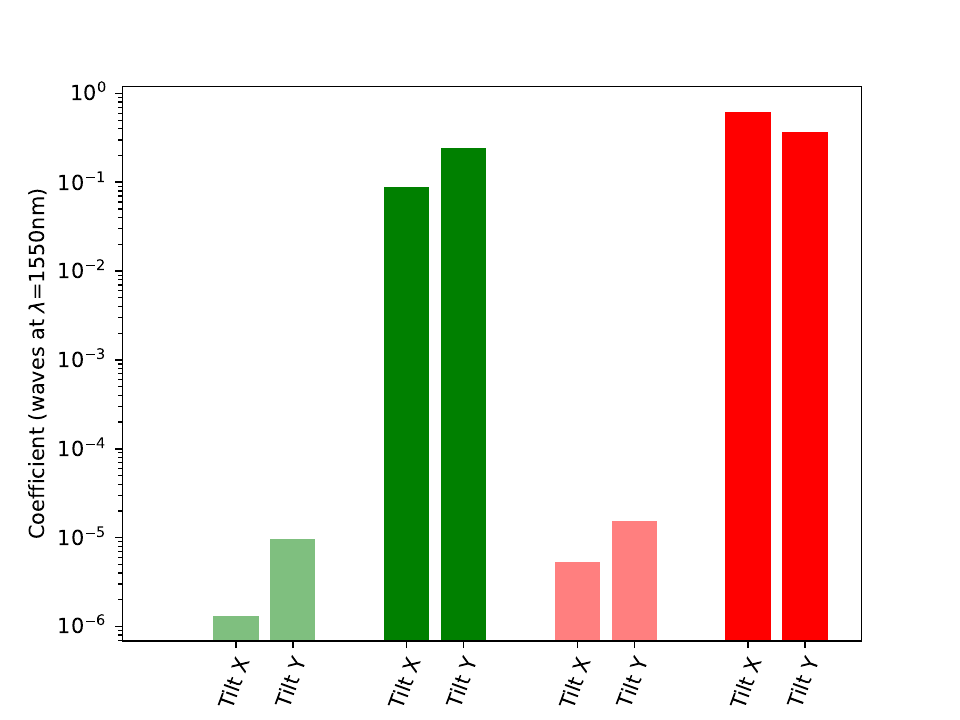}
\includegraphics[width=4.3cm]{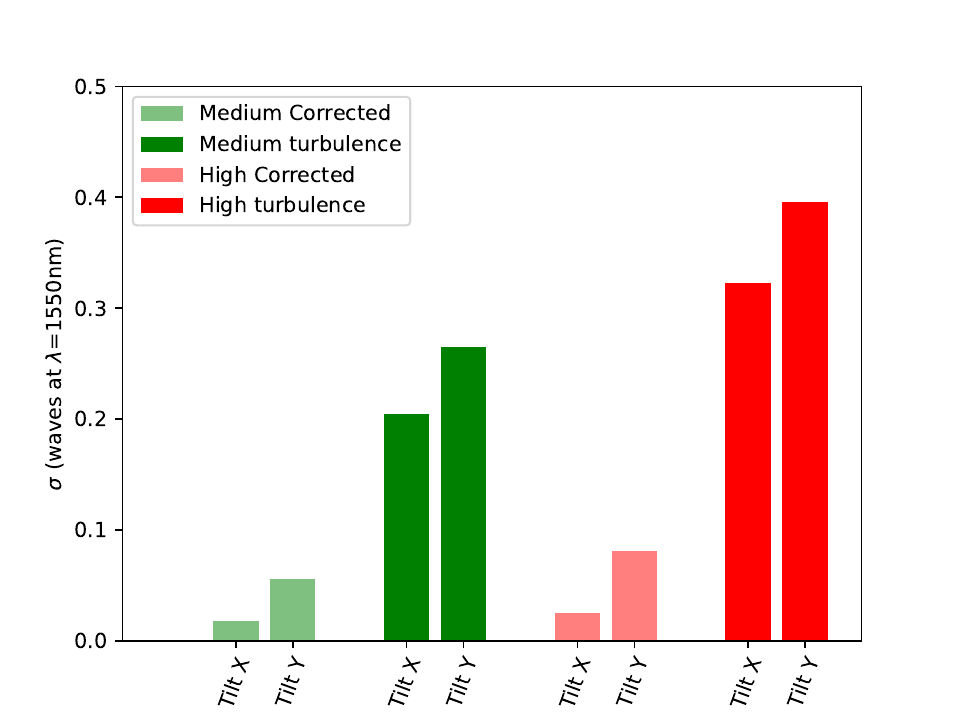}
\caption{\textbf{(left)} Coupling efficiency measured with the system with FSP. Different coupling efficiencies refers to acquisitions with different level of turbulence: medium turbulence not corrected (0.05±0.05), medium turbulence corrected (0.15±0.09), high turbulence not corrected (0.01±0.01), high turbulence corrected (0.07±0.06). \textbf{(center)} Mean tilt values of the acquisitions. Y-axis is logarithmic. \textbf{(right)} Standard deviations of the tilt of the acquisitions.}
\label{fig:resFSP}
\end{figure}

The coupling efficiency with the \gls{fsp} was measured with different levels of turbulence, with medium turbulence corresponding to half strength of the heater (3/5) and high turbulence to full strength (5/5).
Figure~\ref{fig:resFSP} shows the most significative results: the \gls{fsp} is able to reduce both the X and Y tilt standard deviations to less than 0.1 waves for both tested turbulence strengths.
However, this translates to an improvement of the probability distribution of the coupling efficiency only for the medium turbulence, while for the high turbulence both the non-corrected and corrected case have an exponential probability distribution, with higher average value in the latter.
This is due to the presence of higher order aberrations that the \gls{fsp} is not able to compensate, especially in the high turbulence case.

\begin{figure}[htbp]
\centering
\includegraphics[width=4.3cm]{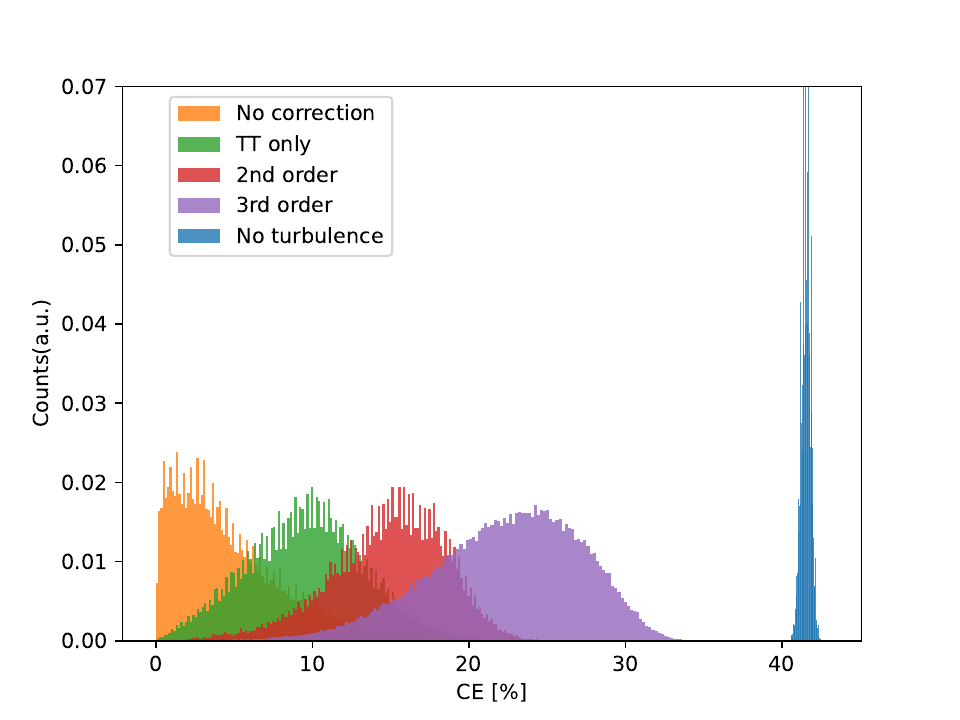}
\includegraphics[width=4.3cm]{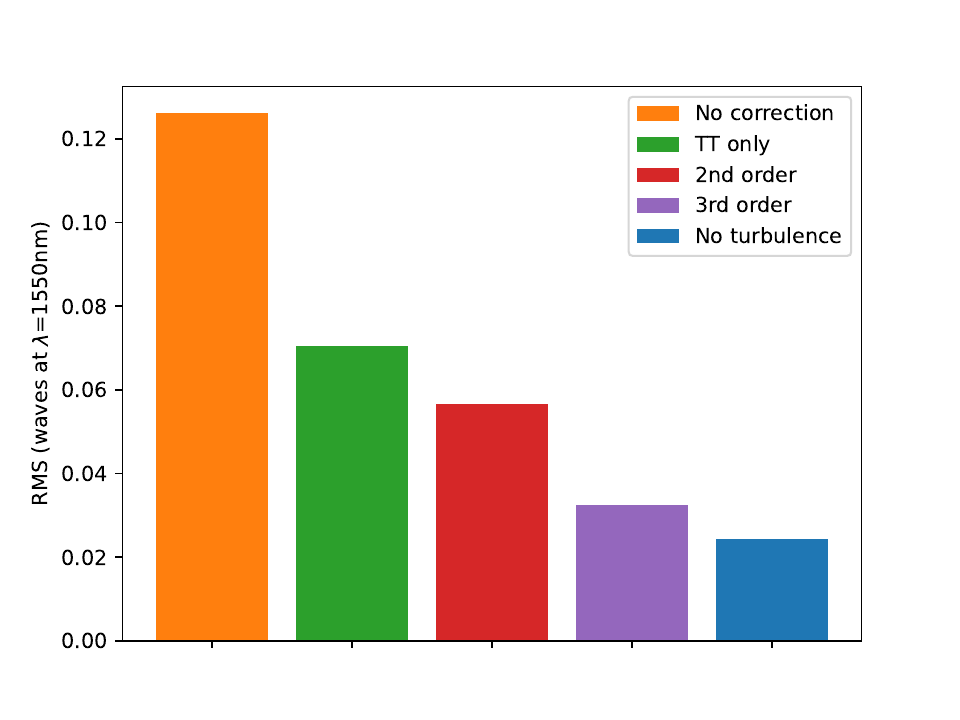}
\includegraphics[width=4.3cm]{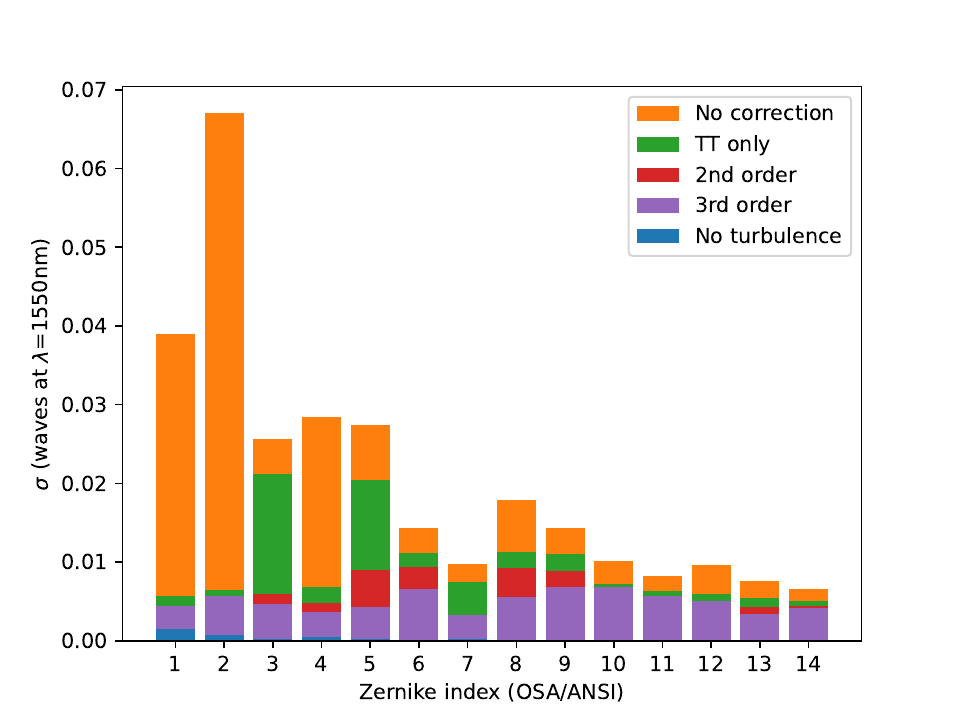}
\caption{\textbf{(left)} Coupling efficiency measured with the system with \gls{mal}. Different coupling efficiencies refers to acquisitions with different levels of correction: no correction (0.05±0.04), correction of tip tilt only (0.10±0.04), correction of Zernike’s modes up to 2nd order (0.15±0.04), correction of modes up to 3rd order (0.22±0.05), closed loop with no turbulence generated (0.415±0.003). \textbf{(center)} Mean RMS of the acquisitions. \textbf{(right)} Standard deviations of the Zernike’s modes of the acquisitions.}
\label{fig:orders}
\end{figure}

The coupling efficiency with the \gls{mal} correction was measured with different levels of correction: tip and tilt modes, modes up to the 1st order, modes up to the third order.
Figure~\ref{fig:orders} shows the results of the experiment for a low-to-medium turbulence level, corresponding to a little less than half strength of the heater (2/5).
As expected, the coupling efficiency increases with the order of modes corrected, while the root mean square error of the phase and the standard deviation of the Zernike's modes that describe the phase at the receiver decrease.
The results show that the correction of aberrations with the \gls{mal} increases the fiber coupling efficiency of single mode fiber injection from about 5\% in absence of correction to more than 20\%.\\

\subsection{Field test}

The field test took place on the 18th and 19th May 2023, two cloudy late spring days characterized by a particularly still air.
The acquisitions were taken during the late afternoon and were characterized by a very mild turbulence, lower than the low-to-medium turbulence level used to test the \gls{mal} in the laboratory.
For this reason, we introduced an artificial turbulence by making a fireplace above the optical path, around 20 meter before the receiver.
This generated a strong turbulence which allowed us to test the operation of the \gls{mal} under different turbulence regimes on a real free-space optical channel.

\begin{figure}[htbp]
    \centering
    \includegraphics[width=4.3cm]{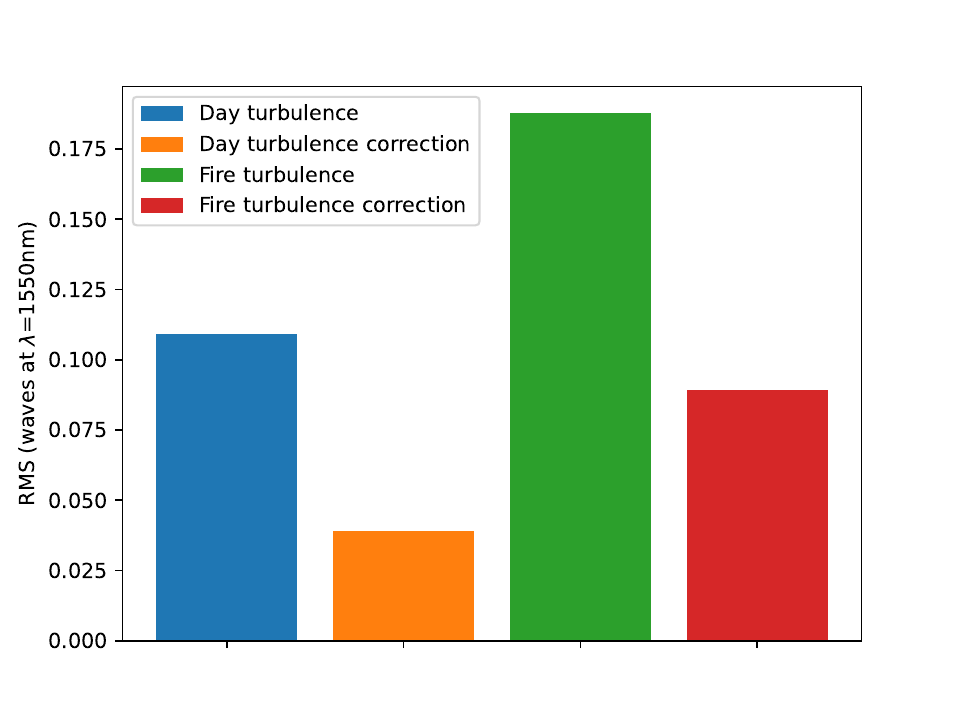}
    \includegraphics[width=4.3cm]{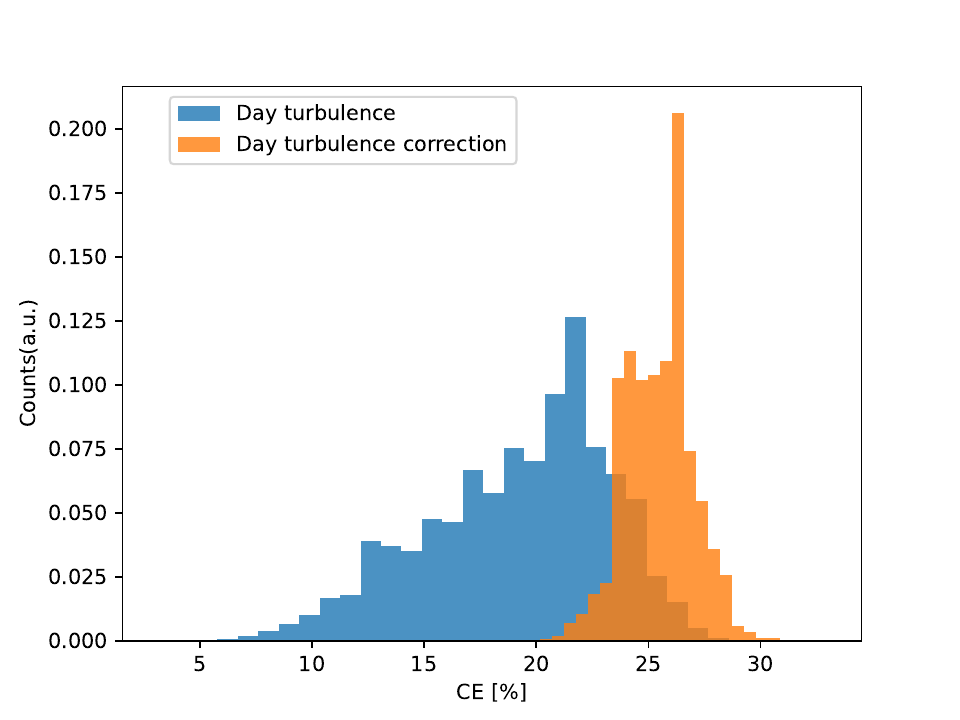}
    \includegraphics[width=4.3cm]{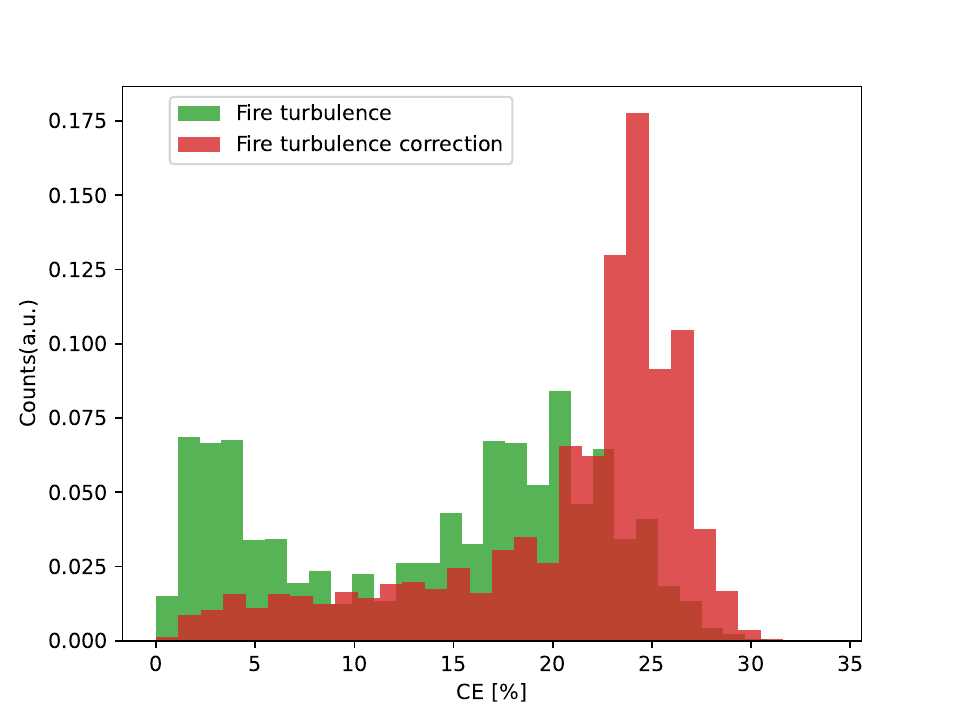}
    \caption{\textbf{(left)} Mean RMS of the acquisitions with the late afternoon turbulence (19/05/2023 at 17:55) and with the fire induced turbulence (19/05/2023 at 18:25). \textbf{(center)} Coupling efficiency with the late afternoon turbulence and no correction ($0.19 \pm 0.04$) or correction up to 3rd order ($0.25 \pm 0.01)$. \textbf{(right)} Coupling efficiency with the fire induced turbulence and no correction ($0.14 \pm 0.08$) or correction up to the 3rd order ($0.21 \pm 0.07$).}
    \label{fig:field}
\end{figure}

The results in Figure~\ref{fig:field} show that the \gls{mal} is able to reduce the root mean square error and increase the coupling efficiency in both turbulence configurations.
Despite the lower improvement in the coupling efficiency with respect to the laboratory experiment, due to the worse condition for the optimization of the coupling efficiency, it is still possible to see a passage from a long tail probability distribution to a Gaussian distribution.
This effect is less evident for the artificial turbulence generated with the fire below the optical path, since the corresponding turbulence is characterized by a stronger weight of the higher orders that the \gls{mal} is not able to correct.
However, the \gls{mal} is still able to increase the mean coupling efficiency and to strongly reduce the effect of fadings in the optical channel, thus providing a clear advantage also in the case of strong turbulence.

\section{Discussion}

The measurements presented in the previous section show that both the \gls{fsp} and the \gls{mal} are effective in reducing the effect of the free-space channel turbulence and consequently increasing the coupling efficiency into \gls{smf} of an optical beam at telecom wavelength.

The \gls{fsp} is effective for correcting the average tip-tilt fluctuations of the optical wavefront for a large range of turbulence.
This, however, brings an improvement in the coupling efficiency only for moderate turbulence levels, due to the presence of higher orders of turbulence that the prism is not able to correct.

This problem is mitigated by the \gls{mal}, which can correct the turbulence up to the 3rd order of Zernike.
This system brings a considerable improvement of the coupling efficiency for moderate turbulence levels both in the laboratory and in a real free-space channel.
A further confirmation of its potentiality comes from its test in a real free-space channel with a fire generated turbulence, where the slight improvement in the coupling efficiency is accompanied by a drastic reduction of channel fadings.

\begin{figure}[htbp]
    \centering
    \includegraphics[width=4.3cm]{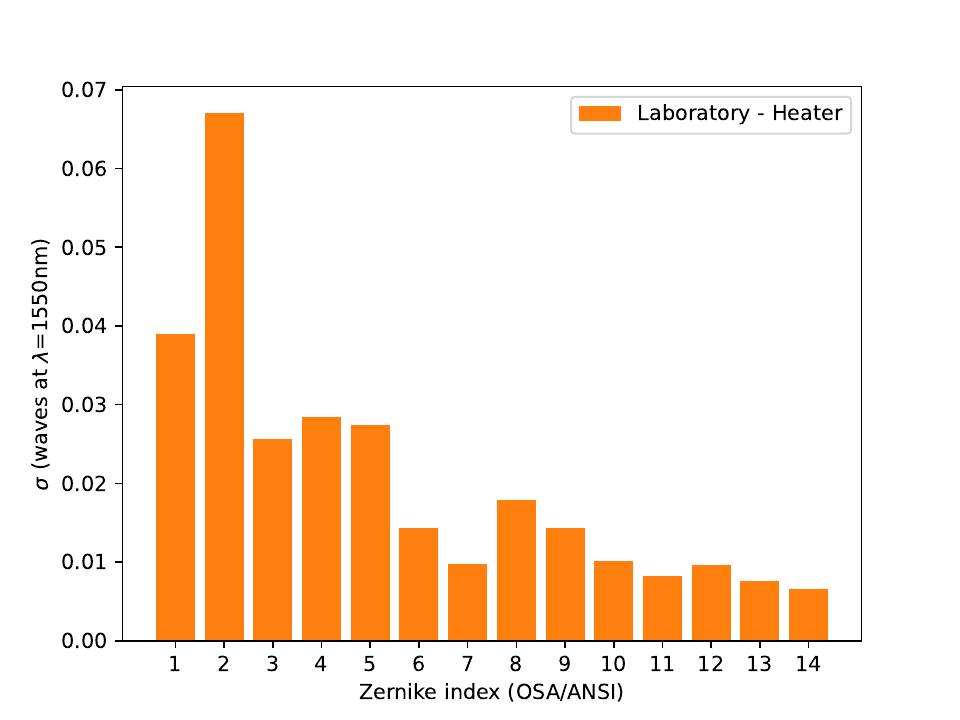}
    \includegraphics[width=4.3cm]{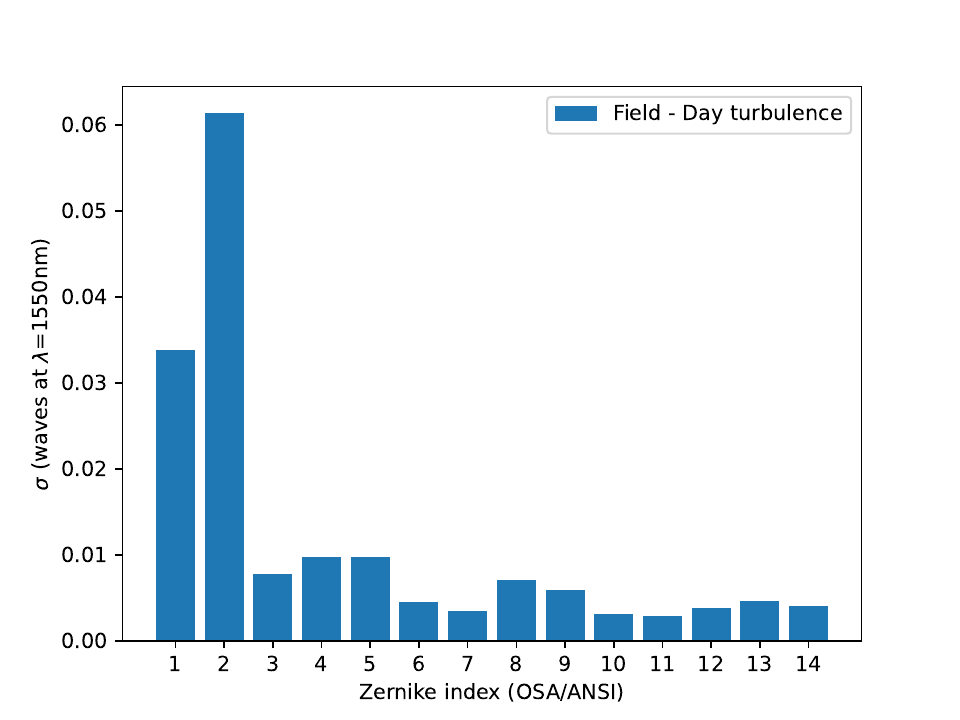}
    \includegraphics[width=4.3cm]{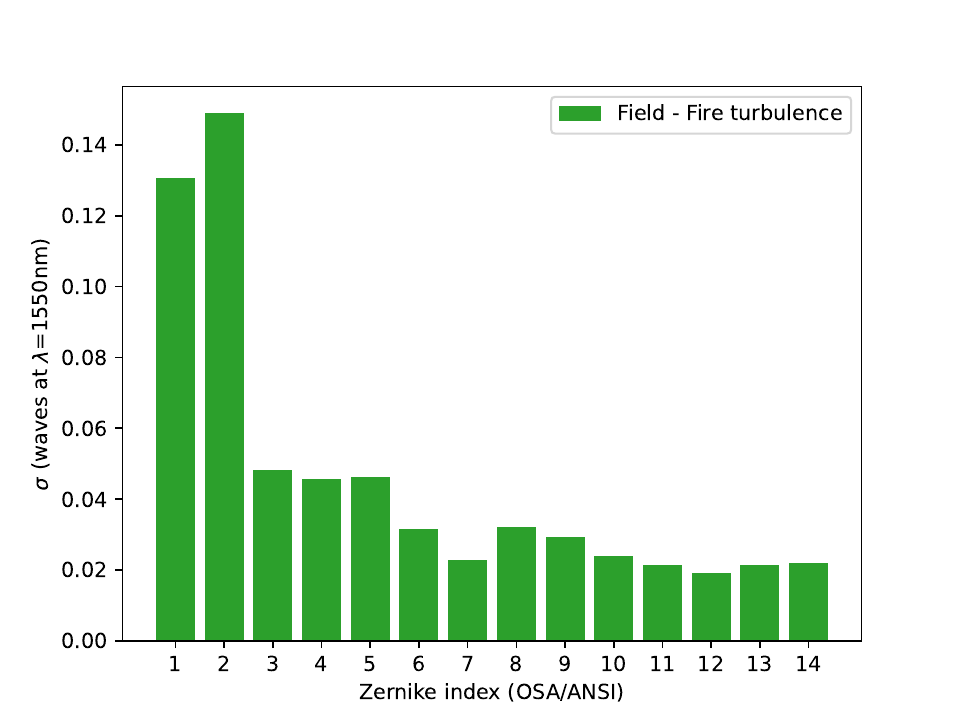}    
    \caption{Standard deviations of the Zernike's modes for the turbulence for \textbf{(left)} the laboratory experiment with the heater at strength 2/5, \textbf{(center)} the free-space experiment and \textbf{(right)} the free-space experiment with a fire under the optical link.}
    \label{fig:turb}
\end{figure}

Figure~\ref{fig:turb} shows the characterization of the channel used for testing the devices through the measurement of the standard deviation of the Zernike's modes.
The turbulence is highly anisotropic, as shown by the different standard deviations of the tip and tilt coefficient.
In the laboratory experiment, the anisotropy is due to the strong temperature difference between the surface of the heater, placed just below the optical channel, and the rest of the room, causing a strong vertical air current.
The free-space experiment took place in a channel located few meters above a grass ground, during a warm cloudy day with slight rain around noon.
These conditions are a possible explanation of the observed anisotropy, due to the vertical air flow caused by the evaporation of the humidity in the grass, combined with an absence of horizontal winds.
The figure shows also how a presence of a source of heat just below the optical channel is accompanied by a strong influence of the higher order Zernike's modes.
Since these channels might not be representative to all the configurations in which such devices might be of interest, a more complete study of their performance would require the implementation of controlled turbulence levels, using a fully characterized turbulence simulator~\cite{Velluet2020}.

\section{Conclusion}

In this work, we show that transmissive elements can be employed for the construction of very compact receiving systems for small aperture telescopes at telecom wavelengths.
The \gls{fsp} has proven to be the most interesting choice if simplicity is required, at the expenses of a lower coupling efficiency, especially for strong turbulence.
The \gls{mal}, on the other hand, has shown good performance in both the laboratory and the free-space channel, with a consistent increase of the coupling efficiency in the case of moderate turbulence and a reduction of the fadings for high turbulence.
This article shows that both devices represents a convenient solution for the implementation of optical communication on short free-space channels.
Their use on longer ground-to-ground channels or for satellite communication requires further investigation and is left for future work.

\bibliography{biblio-main}

\end{document}